\documentclass[reprint,amsmath,amssymb, aps]{revtex4-2}

\usepackage{graphicx}
\usepackage{dcolumn}
\usepackage{bm}
\usepackage{multirow,amsmath,amssymb,xcolor,booktabs,ragged2e} %
\usepackage{here} %

\begin{document}

\title{Nondestructive beam envelope measurements using beam position monitors\\
for low-beta heavy ion beams in superconducting linear accelerator}


\author{Takahiro Nishi$^{1}$}
\email{Contact author: takahiro.nishi@riken.jp}
\author{Tamaki Watanabe$^{1}$}%
\author{Taihei Adachi$^{1}$}%
\author{Ryo Koyama$^{2}$}%
\author{Naruhiko Sakamoto$^{1}$}%
\author{Kazunari Yamada$^{1}$}%
\author{Osamu Kamigaito$^{1}$}%
\affiliation{%
\it {}$^1$ RIKEN Nishina Center for Accelerator-Based Science, 2-1 Hirosawa, Wako, Saitama 351-0198 Japan
}%

\affiliation{%
\it {}$^2$ SHI Accelerator Service Ltd., 7-1-1, Nishigotanda, Shinagawa, Tokyo, 141-0031 Japan
}%
\date{\today}

\begin{abstract}
In superconducting linear accelerators (linacs), accurately monitoring beam dynamics is essential for minimizing beam losses and ensuring stable operations. However, destructive diagnostics must be avoided in superconducting sections to prevent the occurrence of particulates and outgassing, rendering direct measurements of the beam envelope particularly challenging. This study presents a non-destructive method that uses beam position monitors (BPMs) to estimate the transverse beam envelope based on measurements of the quadrupole moment of the beam distribution. Although this concept was originally proposed in the 1980s, its application, especially to hadron beams, has been limited because of low signal sensitivity and the accuracy constraints associated with conventional BPM geometries. To overcome these challenges, we employed $\cos{2\theta}$-type BPMs, which offer improved sensitivity to quadrupole components and are well-suited for low-$\beta$ heavy ion beams. This method was applied to the heavy ion beams in the superconducting RIKEN linac (SRILAC), for which data from eight BPMs were combined with transfer matrix calculations and supplemental wire scanner data. The resulting beam envelope estimates exhibited good agreement with conventional quadrupole scan results, demonstrating the feasibility of this technique for routine, non-destructive beam monitoring in superconducting accelerator sections.
\end{abstract}

\maketitle

\section{Introduction: Beam envelope estimation in superconducting linacs}

In recent years, superconducting linacs have been adopted at high-power heavy-ion accelerator facilities, such as GANIL, FRIB, RAON, HIAF, and RIBF \cite{Spiral2,FRIB,RAON,HIAF, SRILAC}. The implementation of superconducting cavities enables higher acceleration gradients, which in turn facilitates the delivery of high-quality, high-intensity beams for a lower power consumption. However, despite these advantages, the critical challenge of minimizing the beam losses remains. In superconducting cavities, even small beam losses can possibly damage the superconducting cavities, ultimately reducing their operational lifetime.

Despite the importance of controlling the beam losses, directly measuring the beam envelope within superconducting linacs remains a significant challenge. This difficulty arises from the necessity to avoid using destructive diagnostics in the beamline owing to the risk of particulates and outgassing, both of which can induce strong field emission and compromise the cavity performance. One of the most widely used indirect diagnostic techniques is the quadrupole scan method that employs destructive beam profile monitors, such as wire scanners, placed either upstream or downstream from the superconducting cavities \cite{Nishi:HB21-THBC1}. In this method, the beam phase ellipses are inferred by varying the magnetic fields of quadrupole magnets. However, because the optical settings must be altered for each measurement, this method is not suitable for frequent application while the beam is being delivered to users; thus, its utility as a tool for routinely monitoring beam dynamics is limited.

To address this limitation, this study proposes a non-destructive approach for estimating the quadrupole moment of the transverse beam distribution using beam position monitors (BPMs). This concept was originally proposed in the 1980s \cite{Miller} and has since been explored in the context of both electron \cite{Russell, Suwada:JApplPhys, Suwada:PAC03-ROAB012} and hadron beams \cite{Wang:C-ADS,Tajima:IBIC19-TUPP021, Toyama:IPAC2018, Sounas:IPAC2018, Hwang}. Nevertheless, its application to estimating the beam envelope and emittance has remained limited, particularly for hadron beams with relatively low peak currents, primarily because of constraints in the measurement accuracy. One major factor contributing to this limited application is the fact that button-type or stripline-type BPMs, which have geometries that are suboptimal for accurately measuring quadrupole moments, were used in previous studies.

In this work, we adopt $\cos{2\theta}$-type BPMs, which offer superior sensitivity to quadrupole components owing to their geometry \cite{BEPM}. These BPMs also feature relatively large coverage angles for electrodes, providing an enhanced signal strength even for low-$\beta$ heavy ion beams. We implemented this method for heavy ion beams with an intensity of a few particle-$\mu$A and an energy of approximately 4 to 6 MeV/u at the superconducting RIKEN linear accelerator, SRILAC \cite{SRILAC, SHE}. 
By combining the BPM signals with transfer matrix calculations and supplementary data from wire scanners, we successfully estimated the beam envelope without altering the optical system. The results demonstrated good agreement with those obtained by the conventional quadrupole scan method.

The rest of this paper is organized as follows. Section II explains the principles behind measuring beam envelopes using BPMs, Section III presents an analysis of the $\cos 2\theta$-type BPM signals in the low-$\beta$ region, Section IV describes the application of the results to the measurement of the phase ellipse and beam envelope, and Section V summarizes the conclusions of the study and briefly outlines our future work plans.

\begin{figure*}[hbtp]
 \centering
  \includegraphics[width=\textwidth, bb= 0 0 814 187]{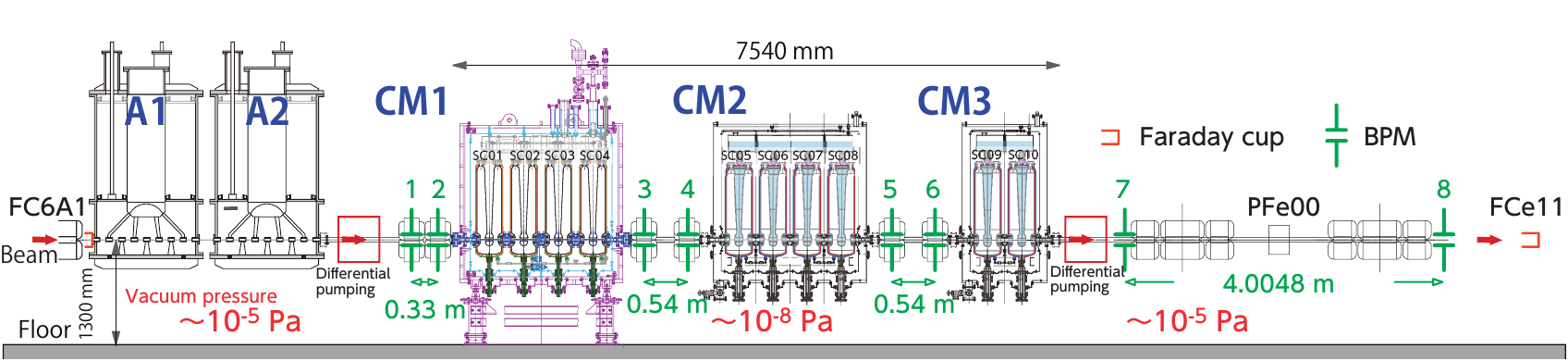}
 \caption{Schematic of the beamline of the SRILAC.
 The green numbers represent our BPMs and ``PFe00'' represents one of the wire scanners. A1 and A2 are normal conducting cavities. CM1--3 are cryomodules that contain superconducting cavities.}
 \label{fig:fig1}
\end{figure*}

\section{Principle of Using BPMs to Measure Beam Envelopes}
To understand the principles underlying the method of measuring the beam envelope using BPMs, we expand the induced voltages across the electrodes ($V_{L, R, U, D}$) using transverse beam multipole moments, such as the dipole ($D_{x,y}$ / $\langle x \rangle$, $\langle y \rangle$), quadrupole ($M_{2}$ / $\langle x^2 \rangle - \langle y^2 \rangle$), and higher-order moments ($M_{ix,y},\ i=3, 4,\cdots$)\cite{Toyama:IPAC2018, Sounas:IPAC2018}, resulting in 
\begin{align}
V_{R} =& I_{\rm beam}(c_0 + c_1D_x + c_2M_2 + c_3M_{3,x}+\cdots)\nonumber \nonumber \\
V_{L} =& I_{\rm beam}(c_0 - c_1D_x + c_2M_2 - c_3M_{3,x}+\cdots)\nonumber \nonumber \\
V_{U} =& I_{\rm beam}(c_0 + c_1D_y - c_2M_2 + c_3M_{3,y}+\cdots)\nonumber \nonumber \\
V_{D} =& I_{\rm beam}(c_0 - c_1D_y - c_2M_2 - c_3M_{3,y}+\cdots),
\label{eq:BPM_signals}
\end{align}
where $I_{\rm beam}$ denotes the beam intensity and $c_{i}$ denotes the coefficient of each multipole term. By neglecting the higher order terms in this equation, $D_{x,y}$ and $M_2$ can be expressed as 
\begin{align}
D_{x,y} =& \langle x,y \rangle \nonumber \\
        =&\ k_{x,y}\frac{V_{R,U} - V_{L,D}}{V_{L} + V_{R} + V_{U} + V_{D}},
\end{align}

and
\begin{align}        
M_2 =& \langle x^2 \rangle - \langle y^2 \rangle \nonumber \\
    =&\ k_q\frac{V_{R} + V_{L} - V_{U} - V_{D}}{V_{L} + V_{R} + V_{U} + V_{D}}, 
	 \label{eq:quadrupole}
\end{align}
respectively, where $k_{x,y}$ and $k_q$ correspond to $c_1/2c_0$ and $c_2/c_0$ in Eq. (\ref{eq:BPM_signals}). For the following analysis, we define $\hat{Q}\equiv M_2 - D_x^2 + D_y^2 = \sigma_x^2 -\sigma_y^2$. These values are derived from the BPM signals via Eq. (\ref{eq:quadrupole}). They can also be obtained using the $\sigma$ matrix at a given point and the transfer matrices between that point and the BPMs. Therefore, based on the calculated transfer matrices, the $\sigma$ matrix and beam envelope are estimated by fitting the $\sigma$ matrix elements to reproduce the measured $\hat{Q}$.

For this estimation to be valid, the beam loss must be negligibly small. Using the Faraday cups (FC) positioned upstream (6A1) and downstream (e11) of the SRILAC, which are shown in Fig.~\ref{fig:fig1}, the transmission efficiency was confirmed to exceed 95\%, ensuring that this condition was satisfied.

\section{Analysis of $\bm{\cos{2\theta}}$-type BPM Signal in Low-$\bm{\beta}$ Region}
This section discusses the $\cos 2\theta$-type BPM used in this study, including its advantages for conducting second-moment measurements and the disadvantage of its bias, which is present primarily when measuring low-$\beta$ beams. We demonstrate that this disadvantage can be mitigated by performing a double integration of the signals.

\begin{figure}[t]
  \centering
  \includegraphics[width=0.5\textwidth, bb = 0 0 1024 768]{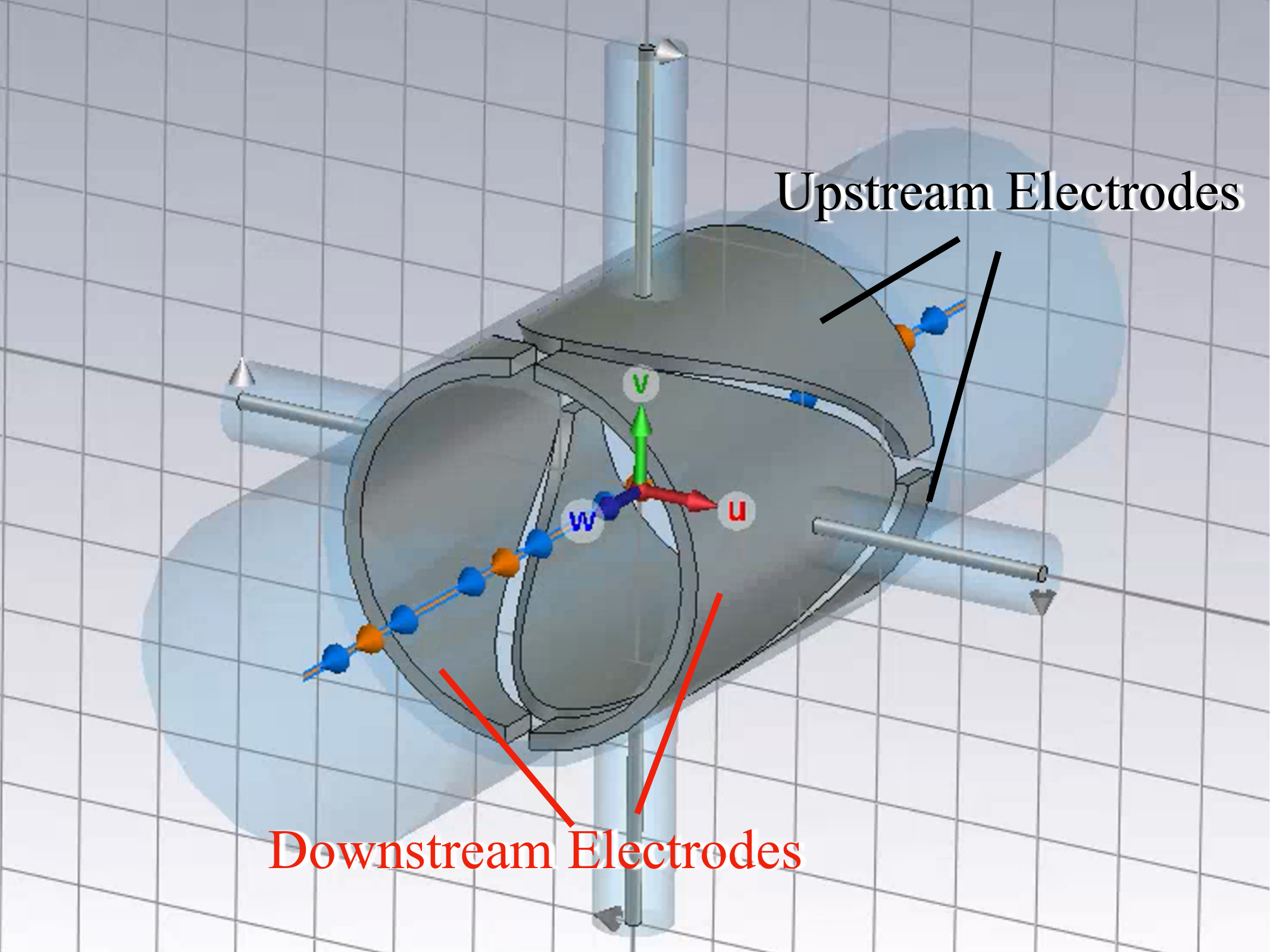}
  \caption{3D model of a type-A BPM constructed using CST Studio Suite. The beam is directed toward the lower left corner. The upper/lower and right/left electrodes represent the upstream and downstream, respectively. 
 }
 \label{fig:cad}
 \end{figure}

Figure~\ref{fig:fig1} shows the layout of the SRILAC, with its beamline and eight BPMs. These BPMs measure not only the beam positions but also the beam energy in each section in between the superconducting cavities; thus, we also call the BPMs beam energy position monitors (BEPMs). 
Two types of BPMs are used: type-A (numbers 1 to 6) and type-B (numbers 7 and 8). 
The parameters of these BPMs are summarized in Table~\ref{tab:BEPM}. Figure~\ref{fig:cad} shows the 3D model of a type-A BPM constructed using CST Studio Suite. As the figure shows, our BPMs have $\cos2\theta$-shaped electrodes. 

\begin{table}[b]
\caption{\label{tab:BEPM}%
Parameters of the type-A and type-B BPMs.}
\begin{ruledtabular}
\begin{tabular}{ccccc}
 & \textbf{longitudinal} & \textbf{inner}  & $\mathbf{k_{x,y}}$ & $\mathbf{k_q}$\\
  & \textbf{length} & \textbf{radius}  &  & \\
 
\hline
         type A   & 50 mm & 20 mm &  26 mm& 570 mm$^2$  \\
         type B   & 60 mm & 30 mm &  39 mm& 1200 mm$^2$ \\         
\end{tabular}
\end{ruledtabular}
\end{table}

These $\cos2\theta$-shaped electrodes enabled us to precisely measure $\hat{Q}$. Following the conventions outlined in Ref. \cite{Miller}, the characteristics of this type of BPM are described as follows. First, given a delta-function line current $I_{\rm beam}$ at the point ($r$, $\phi$), the image current density $J_{\rm image}$ on a conducting cylinder of radius $a$ at the position ($a$, $\theta$) can be expressed as a series expansion of powers of ($r/a$):

\begin{align}
J_{\rm image}(r,\theta, \phi,a)= \frac{I_{\rm beam}(r,\phi)}{2\pi a}\sum_{k=0}^\infty \left(\frac{r}{a}\right)^{k}\cos{k(\theta - \phi)}.
\label{eq:BPM_signals}
\end{align}
The image charge induced on each electrode is given by the convolution of $J_{\rm image}$ with the electrode's geometrical shape. Therefore, for measuring the zeroth moment (i.e., the beam current itself), an undivided electrode configuration is ideal \cite{Koyama:PASJ2024}. For the dipole moment, which corresponds to the beam position, diagonally cut (i.e., $\cos{\theta}$-shaped) electrodes are optimal, although each electrode pair is sensitive only to either $D_x$ or $D_y$. 
Similarly, the quadrupole moment is ideally measured using two pairs of $\cos{2\theta}$-shaped electrodes. 
In the case of our BPMs, a pair of electrodes (left and right or up and down) forms a single $\cos{2\theta}$-shaped electrode, as illustrated in Fig.~\ref{fig:cad}.
Consequently, the coefficients $c_0$, $c_1$, and $c_n$ ($n \geq 3$) in Eq. (\ref{eq:BPM_signals}) ideally vanish. In contrast, for typical stripline- or button-type BPMs, these coefficients are nonzero, necessitating higher-order corrections and resulting in lower precision.

Deriving $\hat{Q} \equiv \sigma_x^2 - \sigma_y^2$ requires obtaining the beam position as well as the quadrupole moments. Although the $\cos{2\theta}$-shaped electrodes are not ideal for determining the beam position, precise calibration using a wire allowed us to estimate the beam position from the four electrode signals at an accuracy of approximately 100 $\mu$m \cite{BEPM:Calib}. By integrating these measurements, the $\cos{2\theta}$-type BPM was used to determine $\hat{Q}$ with high precision.

Although the $\cos{2\theta}$-type BPM offered this advantage, its structural design introduces a bias in the electrode signals. To understand this effect more comprehensively, we formulated the signals obtained from each electrode using a simplified equivalent circuit model. In a cylindrical BPM, for a given electrode coverage angle $\phi$, the time-dependent signal voltage $V$ from that electrode is expressed as 
\small
\begin{eqnarray}
V(\phi,t) ={\rm Re}\left[\Sigma_{n=0}^{\infty}\frac{jn\omega_{RF} R}{1 + jn\omega_{RF} RC}\frac{\phi LI_0}{\pi\beta c}\right. \nonumber\\
\left. \times\exp{\left(\frac{n^2\omega_{RF}^2\sigma_t^2}{2} + jn\omega_{RF} t\right)}\right],
\end{eqnarray}
\normalsize
where $R$ and $C$ represent the resistance and capacitance of the equivalent circuit of the electrodes, respectively; $L$ represents the longitudinal length of the electrodes; $t$ denotes the time; and $\beta$, $\sigma_t$, and $\omega_{RF}$ denote the beam velocity, time width, and angular velocity of the 73-MHz radio frequency (RF) field, respectively. To account for the shape of the electrode, it is divided into segments of $\delta l$ (incorporating the l-dependence of $\phi$) via
\begin{equation}
V(t) = \frac{1}{L}\int^{L/2}_{-L/2}\left\{\frac{\phi(l)}{\bar{\phi}}V\left(t-l/\beta c\right)\right\}dl,
\label{eq:BPM_signals_geo}
\end{equation}
where $\bar{\phi}$ represents the electrode's averaged coverage angle, $\pi/2$. For $\cos2\theta$-type BPMs, upstream and downstream electrodes have coverage angles of $\phi(l) = \arccos(2l/L)$ and $\arccos(-2l/L)$, respectively.

\begin{figure}[t]
    \centering
\includegraphics[width=0.5\textwidth]{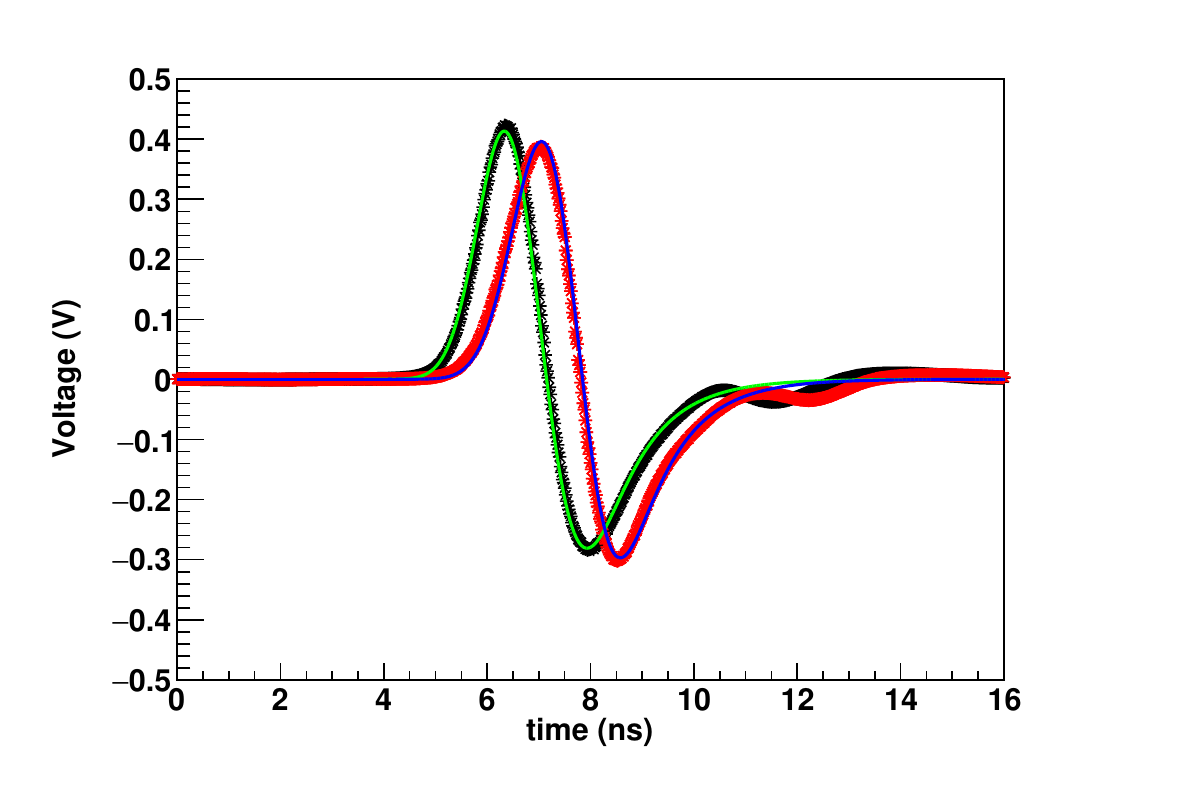}
  \caption{Waveforms of the averaged signals from the BPM electrodes. The black and red points represent the data from the upstream and downstream electrodes, respectively, and the green and blue lines represent the corresponding fitting results obtained using Eq. (\ref{eq:BPM_signals_geo}).}
    \label{fig:waveform}
\end{figure}

Next, we compared the averaged signals acquired from the upstream and downstream electrodes of BPM6 after amplification by a 36-dB amplifier and digitization via a waveform digitizer, with the theoretical fits based on the aforementioned model, as shown in Fig.~\ref{fig:waveform}. The value of $\hat{Q}$ at BPM6 was estimated to be sufficiently small ($<$ 1 mm$^2$) based on the quadrupole scan measurement; this allowed the theoretical fit functions to be compared without being affected by a finite transverse beam size. In this analysis, the resistance $R$ was fixed as the measured value (50 $\Omega$), whereas $\sigma_t$, $C$, the signal amplitude, and the timing were treated as free parameters. These fitting parameters were commonly used for analyzing the signals from the upstream and downstream electrodes. The relative peak positions and heights were not treated as variables but were derived based on the geometry of the electrodes. As indicated by the figure, even this simplified model successfully reproduced the experimentally observed waveforms.

Most importantly, both the experimental data and theoretical model indicated that the signal obtained from the upstream electrodes was larger than that from the downstream electrodes. The difference was primarily attributed to the asymmetry of the electrode geometry relative to the beam direction. In the upstream electrode, a large signal corresponding to a large $\phi$ was generated, followed by gradually overlapping signals corresponding to a smaller $\phi$ with a time delay. The peak maximum was primarily determined by the signal from the upstream portion of the electrode that exhibited a larger $\phi$. In contrast, in the downstream electrode, a relatively small signal corresponding to a small $\phi$ was generated, followed by gradually overlapping signals corresponding to a larger $\phi$ with a time delay. In the latter case, although the peak maximum was generated by the portion of the downstream electrode that exhibited a larger $\phi$, the undershoot of the signal from the upstream portion overlapped with the peak maximum, resulting in a relatively small peak compared to that of the upstream electrode.
This effect is expected to be significant when the beam bunch length is comparable to or shorter than the longitudinal length of the electrode. Considering realistic beam bunch lengths, this effect is particularly pronounced for low-$\beta$ particles. According to the simulation performed using CST Studio Suite, for a beam bunch length of $\sigma_t$ = 0.5~ns (corresponding to the fitted values of the experimental data), the deviation of the signal strength was minimal (0.2\%) at $\beta$ = 0.99. However, in our case ($\beta$ = 0.1), the deviation was non-negligible, reaching approximately 5\%.

One possible solution to this is evaluating and correcting this bias effect. In Ref.~\cite{Nishi:SRF2023}, we demonstrated the validity of this method using one day's worth of data. However, this method requires estimating the correction factor from experimental data in advance. Consequently, this may cause systematic errors (depending on the beam conditions under which the data are collected) because the correction factor is affected by the longitudinal distribution of the beams.

\begin{figure}[t]
  \centering
    \includegraphics[width=0.5\textwidth,bb= 0 0 1024 768]{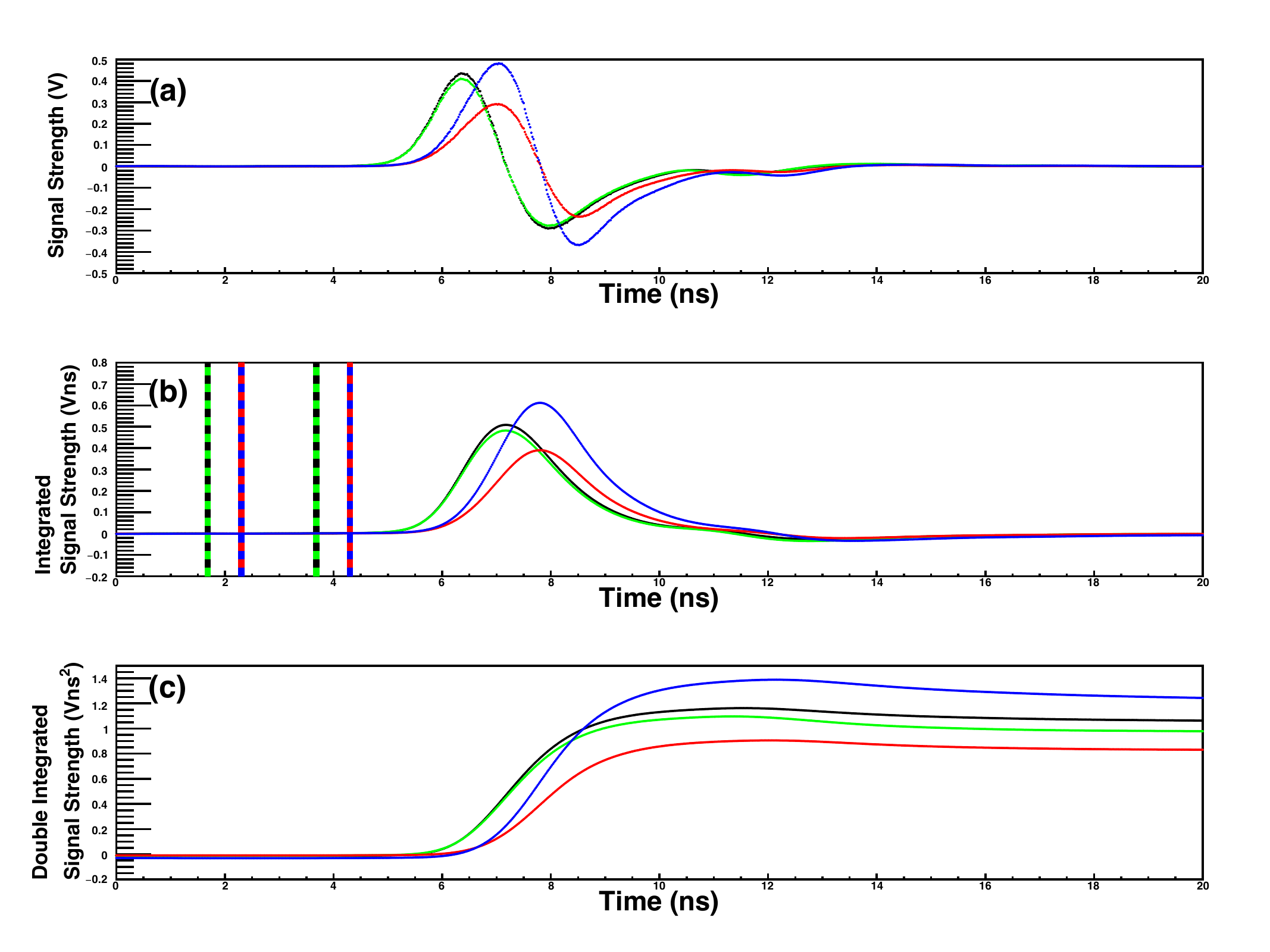}
  \caption{(a) Raw BPM waveform signals from the right (black), left (green), upper (red), and lower (blue) electrodes. (b) Integrated waveform signals obtained after the slope corrections were performed. The vertical lines correspond to the regions used to calibrate the background level (see the text for details). (c) Double-integrated waveform signals. }
  \label{fig:Double_integral}
\end{figure}

After a detailed investigation, we found an alternative solution that utilizes double-integrated signals. As explained above, the bias effect is caused by the time differences between the signals originating from different positions on the electrodes. Hence, integrating the signals cancels out this effect. An analysis based on the simulation indicated that the bias effect can be reduced to 1\% or less if the signals are integrated and eliminated for any beam bunch with double-integrated signals \cite{Adachi:PASJ2023}. Consequently, we adopted the double integration method to process the experimental data. Figure~\ref{fig:Double_integral} shows the raw, integrated, and double-integrated waveform signals from a type-A BPM. To perform the integration, the background slope caused by the offset of the raw waveform signals was subtracted from the integrated waveform signals. This slope satisfied the cyclic boundary conditions and ensured that the integrated signal was zero in regions without signals (4.5 $\pm\ 1$ ns before the peak for the type-A BPMs and 5.5 $\pm\ 1$ ns for the type-B BPMs; these are shown as vertical lines in the figure). 
Because the reflected signals followed the obtained signals, we used the maximum of the double-integrated signals to calculate $\hat{Q}$. 

\begin{figure}[t]
 \centering
 \includegraphics[width=0.5\textwidth,bb=0 0 743 617]{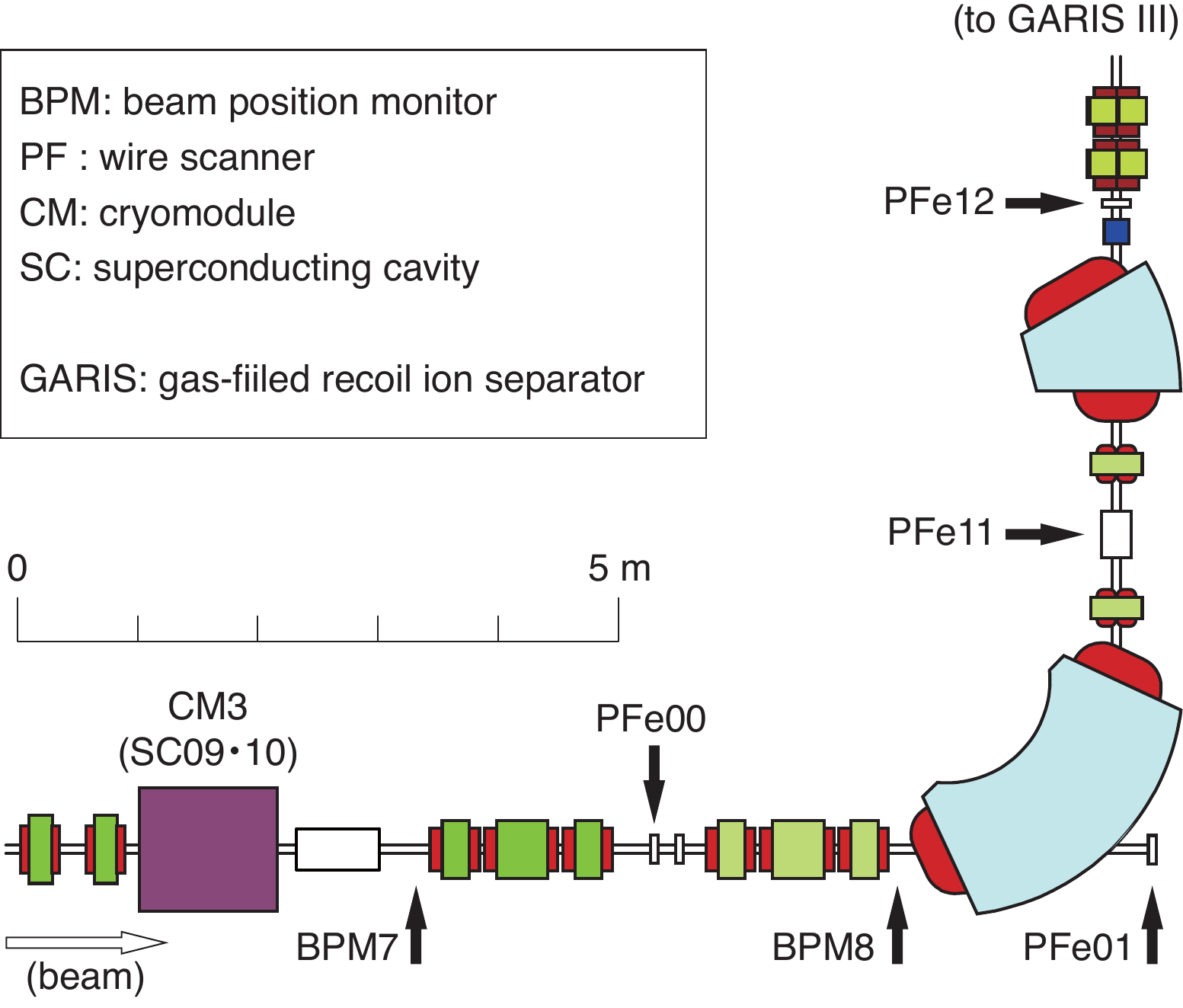}
 \caption{Top view of the beamline from the SRILAC to the experimental target. ``PF'' denotes the wire scanners used to measure the beam profiles.}
 \label{fig:topview}
\end{figure}

\section{Application to Measurements of the Phase Ellipse and Beam Envelope}
We now present the practical application of the methods described above. To evaluate the accuracy of the $\hat{Q}$ values measured by the BPMs, we compared them to the four datasets of conventional quadrupole scan (Q-scan) measurements conducted within the past year. Typically, the beam phase ellipse is measured at the wire scanners (the so-called profile monitors, or PFs) located downstream of the triplet quadrupole magnets at PFe00 and PFe01, as shown in Fig.~\ref{fig:topview}, after reducing the beam intensity to $\simeq$ 100 electric nA. 
The transfer matrices from each BPM to e00 were calculated using the beam dynamics simulations that included the effects of acceleration, 
which predicted the phases with an accuracy of 10$^\circ$. 
Using the measured phase ellipse and the transfer matrices, the expected $\hat{Q}$ values at each BPM were calculated and compared to those measured by the BPMs when the beam intensity was maintained at a few tens of electric $\mu$A.
Based on this comparison, we applied correction offsets of up to 10 $\mathrm{mm}^2$ to the values measured by the BPMs to align these values with the results of the Q-scan method. After this correction, the distribution of the differences between the corrected $\hat{Q}$ values and the $\hat{Q}$ values derived by the Q-scan method was used to estimate the $\hat{Q}$ measurement errors at the BPMs. We assumed that the Q-scan method provided sufficiently accurate values and that the $\hat{Q}$ measurement errors (denoted as $\sigma_{\hat{Q}}$) were uniform across all BPMs. Under this assumption, we estimated $\sigma_{\hat{Q}}$ using the sample variance of the differences between the values measured by the eight BPMs, requiring these variances to follow a $\chi^2$ distribution. The value of $\sigma_{\hat{Q}}$ was estimated to be approximately $4.5 \pm 1\ \mathrm{mm}^2$, while the typical values of $\hat{Q}$ ranged from $-30$ to $30$ $\mathrm{mm}^2$.

\begin{figure*}[bhtp]
 \centering
 \includegraphics[width=\textwidth,bb=0 0 1920 1080]{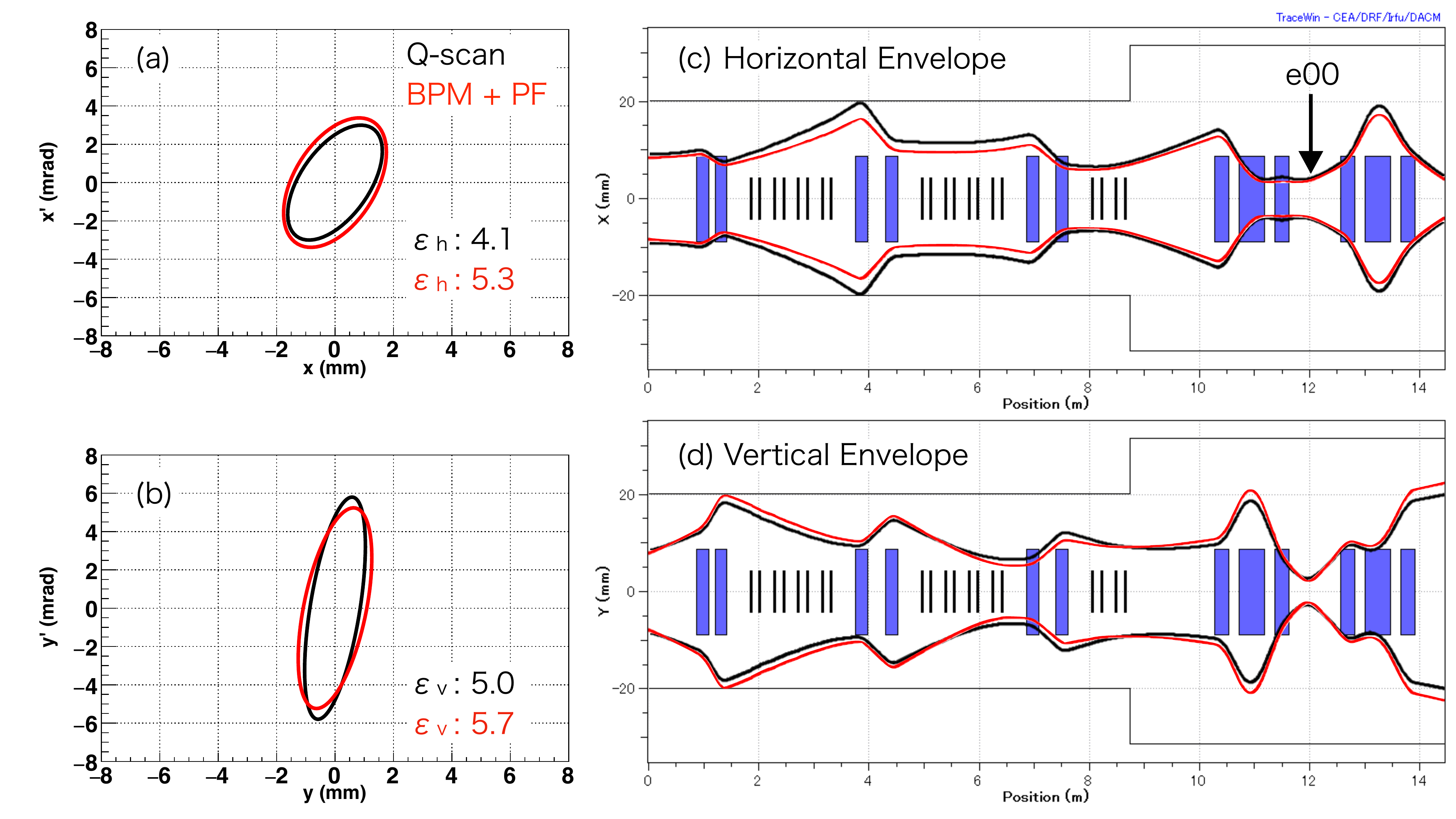}
 \caption{
(a) Horizontal and (b) vertical phase ellipses at e00 (shown in Fig.~\ref{fig:topview}) estimated using the conventional Q-scan method (black) and our method employing BPM signals with PF data (red). \newline (c) Horizontal and (d) vertical 5-$\sigma$ SRILAC beam envelopes generated by the beam dynamics simulation software TraceWin and based on the transfer matrices and estimated phase ellipses calculated by the conventional Q-scan method (black) and our method employing BPM signals with PF data (red). The blue and black rectangles represent the quadrupole magnets and RF cavities, respectively. The solid lines correspond to the beam duct.}
 \label{fig:fig_Envelope}
\end{figure*}

\begin{table*}[hbtp]
\caption{\label{tab:fit_result}%
Phase ellipse parameters at e00 and $\chi^2/n.d.f.$ for each estimation method. In the fourth column, the absolute emittance values are constrained to within 2 $\sigma$ of the BPM + PF result. The value marked with an asterisk ($^{\ast}$) is at the boundary of the constraints.}
\begin{ruledtabular}
\begin{tabular}{cccc}
\textbf{Phase ellipse } & \textbf{Q-Scan} & \textbf{BPM + PF}  & \textbf{BPM + $\epsilon$ constraints}\\
\textbf{parameters at e00} &  &   &  \\
\hline
         {$\epsilon_h$}    & 4.1 $\pm$ 0.3 & 5.3 $\pm$ 0.8 &  5.3           \\
         {$\alpha_h$}      & -0.66         &          -0.53&  -0.56\\         
         {$\beta_h$}       & 0.65          &           0.59&   0.45\\         
         {$\epsilon_v$}    & 5.0 $\pm$ 0.4 & 5.7 $\pm$ 0.6 &  4.5$^{\ast}$\\
         {$\alpha_v$}      & -0.68         &          -0.58&   0.17\\         
         {$\beta_v$}       & 0.22          &           0.28&   0.15\\         
        \midrule		         
         {$\chi^2/n.d.f.$} & 35.3/16       &         15.7/7&    3.5/2\\         %
\end{tabular}
\end{ruledtabular}
\end{table*}

Using the corrected $\hat{Q}$ values and the transfer matrices, the phase ellipse was reconstructed. However, a data analysis indicated that the uncertainties in the absolute values of the beam emittances were unacceptable. This occurred because $\hat{Q}$ is sensitive to the relative values of $\epsilon_h$ and $\epsilon_v$, but is not sufficiently sensitive to their absolute values. To address this issue, we developed a method that uses additional information from the PFs. To increase the sensitivity to the absolute beam size and render the method more practical, we utilized the $\sigma_x$ and $\sigma_y$ values measured by the PFs located at e00, e11, and e12, which are shown in Fig.~\ref{fig:topview}. This additional information on the absolute beam size enhanced the sensitivity to the absolute beam emittance values. Although this method required a destructive detector, it did not necessitate changes in the strengths of the quadrupole magnets, unlike the Q-scan method.

Figures~\ref{fig:fig_Envelope} (a) and (b) compare the phase ellipses reconstructed by the conventional Q-scan method (black) and our new method of using BPM data and PFs (red), respectively. As shown in the figures, we succeeded in reconstructing not only the shapes of the phase ellipses but also the absolute emittances, thus achieving agreement with the conventional method within an accuracy of 30\%.
If we used only the BPM data, the estimated vertical emittance dropped to almost zero.
Using the reconstructed phase ellipse, we calculated the beam envelopes via the beam dynamics simulation software TraceWin, as depicted in Figs.~\ref{fig:fig_Envelope} (c) and (d). The beam envelope estimated according to the analysis of the BPM and PF data was remarkably similar to that obtained by the Q-scan method. For example, the two results agreed to within an accuracy of 10--20\% at the point where the beam duct and beam envelope are closest. These estimated envelopes are valuable for deducing the beam loss and monitoring changes in the beam conditions. Although this method is technically not a non-destructive beam monitor, it requires only one-shot profiles with the current optics settings, rendering it suitable as a daily ``semi-non-destructive'' beam envelope monitor. In contrast, the conventional Q-scan method requires the optics settings to be changed multiple times, which is not feasible when the beam is being supplied to users. 

For the case in which the beam optics are frequently modified (e.g., during the optimization of the SRILAC optical system), measuring the beam profile with PFs each time may not be practical, as the beam intensity must be reduced to prevent the wires from melting. Even for such cases, the phase ellipse can be appropriately estimated using only the BPM signals (as ``true'' non-destructive monitors) by applying constraints on the absolute emittance values derived from the BPM and PF data or Q-scan results at the start of the optimization process. When optimizing the beam transport within the superconducting cavities, the parameters are roughly adjusted to match the upstream and downstream FC readings to within a few percent. During subsequent adjustments, the beam emittance is assumed to remain constant as long as the transmission efficiency does not significantly change. The phase ellipse parameters estimated by these methods are summarized in Table~\ref{tab:fit_result}. 
In the table, the second and third columns correspond to the results of the Q-scan method and those of our method of utilizing the BPMs and PFs, respectively.
The fourth column presents the results obtained when using only the BPM data constrained by the emittances. In that case, we set the constraints to maintain the beam emittances to within 2$\sigma$ of the BPM + PF result. As shown in the table, the results obtained from each method were qualitatively similar. Consequently, the beam envelope could be visualized instantaneously as the cavity voltage, phase, or quadrupole magnet currents were adjusted.

Because this method relies on BPM measurements of the quadrupole moments of the beam distribution, its sensitivity to the detailed shape of the distribution (particularly in the tails) is limited. Therefore, to ensure a comprehensive understanding of the beam transport dynamics in practical applications, it is essential to use this approach as a complement to other diagnostic tools such as beam loss monitors, X-ray monitors, or vacuum pressure measurements.

\section{CONCLUSION and future perspectives}
In this study, we developed an improved method for estimating the transverse beam envelope in superconducting linacs that utilize BPM signals. Although this method has been recognized for several decades, in only a few instances it has been applied to estimating hadron beam envelopes or beam emittances, primarily because of accuracy issues.
To address this challenge, we devised a method that employs $\cos{2\theta}$-type BPMs, which offer higher sensitivity to the beam size, particularly for low-$\beta$ heavy ion beams with relatively low peak currents. During this process, we discovered that bias effects (arising from short bunch lengths, low-$\beta$ particles, and the $\cos{2\theta}$-shaped electrodes of the BPMs) are crucial for accurately estimating the value of $\sigma_x^2 - \sigma_y^2$ for the beams. By utilizing double-integrated signals and incorporating additional information from wire scanners, we confirmed that it is possible to estimate the beam envelope within a superconducting linac as the beam is being delivered without altering the optical system. This work represents a significant step forward in the practical application of BPM-based estimation of the beam envelope and its integration into routine accelerator operations. Furthermore, we proposed a hybrid monitoring approach that combines non-destructive BPMs with a limited utilization of destructive detectors, enabling continuous beam diagnostics. This strategy offers a promising approach for high-intensity beam accelerator facilities, especially those that employ superconducting cavities for which the implementation of destructive diagnostics poses significant challenges.

In future work, we will refine this method by incorporating additional data from other diagnostic techniques into routine operations to enhance the beam tuning precision. We will also investigate the feasibility of estimating the beam envelope using only BPMs and an appropriately designed lattice, potentially eliminating the need for destructive devices altogether.

\section{ACKNOWLEDGMENTS}
We thank Prof. Toyama for the discussions about the method involving the BPMs, and our colleague Dr. Fukunishi for the discussions in the early stages of development. We are also grateful to all RILAC operations staff, who helped us perform a wide range of measurements for the study. Finally, we acknowledge the use of ChatGPT for its language assistance in improving the clarity and readability of this paper. 

\section{AUTHOR CONTRIBUTIONS}
T.~Nishi contributed to the conceptualization, methodology, formal analysis, software development, simulation, and original draft writing. 
T.~Watanabe contributed to the conceptualization, data curation, hardware development (construction, calibration and operation of the BPMs), software development, and review \& editing the manuscript. 
T.~Adachi contributed to methodology, formal analysis, simulation, and review \& editing the manuscript. 
R.~Koyama contributed to data curation and software development. 
N.~Sakamoto and K.~Yamada contributed to operation of SRILAC and software development, particularly for the beam dynamics in RF cavities. 
O.~Kamigaito contributed through supervision of the project and review \& editing the manuscript.

\bibliography{basename of .bib file}
\end{document}